\documentclass[preprint,groupedaddress]{revtex4}
\usepackage{amsmath, amsfonts, amsthm, amssymb,epsfig}

\def\bib{B\kern-.05em{I}\kern-.025em{B}\kern-.08em}
\def\btex{B\kern-.05em{I}\kern-.025em{B}\kern-.08em\TeX}
\usepackage{graphicx}
\usepackage{epstopdf}

\usepackage[utf8]{inputenc}

\begin{document}


\markboth{G. Uribarri, and G. B. Mindlin}{Heterogeneously forced population of excitatory Adler neurons.}

\title{Resonant features in a forced population of excitatory neurons.}

\author{Gonzalo Uribarri}
\affiliation{Departamento de Física, Facultad de Ciencias Exactas y Naturales, Universidad de Buenos Aires and IFIBA-CONICET}
\email[E-mail me at: ]{gonzalo@df.uba.ar}

\author{Gabriel B. Mindlin}
\affiliation{Departamento de Física, Facultad de Ciencias Exactas y Naturales, Universidad de Buenos Aires and IFIBA-CONICET}


\begin{abstract}
	In recent years, the study of coupled excitable oscillators has largely benefited from a new analytical technique developed by Ott and Antonsen \cite{Ott2008}. This technique allows to express the dynamics of certain macroscopic observable in the ensemble in terms of a reduced set of ordinary differential equations. This makes it possible to build low-dimensional models for the global activity of neural systems from \emph{first principles}. We investigated the macroscopic response of a large set of excitatory neurons to different forcing strategies. We report resonant behavior, that depends on the heterogeneity between the units and their coupling strength. This contrasts with the type of response that an external forcing can elicit in simple and widely used phenomenological models. 

\end{abstract}

\maketitle

\section{INTRODUCTION}
\label{sec:1}
The basic building blocks of a nervous system are its neurons, cells whose dynamical behavior is typically excitable. In many cases, the ultimate output of a nervous system involves the control of a macroscopic biomechanical device, acting as an intermediary between an organism and its environment. This problem can be framed within the open question of how to build a statistical dynamics for out-of-equilibrium ensembles of units. Until recently, a frequent approach consisted in building phenomenological models. In the last years, diverse methods allowed deriving the dynamics obeyed by variables describing large ensembles of excitable units \cite{Ott2008,pikovsky2008,Ott2009,Montbrio2015,Stankovski2017}. Since this breakthrough, dynamicists were able to revisit the $N \to \infty$ limit of many out-of-equilibrium problems \cite{abrams2008,childs2008,laing2009,martens2009,Alonso2011,wolfrum2011,pikovsky2011,laing2014,So2014,Ullner2016,Rodrigues2016}.

One of the simplest neural architectures is a set of excitatory units, with every unit connected to all the others. The average activity of such a configuration has been very successfully modeled by a variable whose dynamics is ruled by a differential equation with a sigmoidal vector field \cite{Wilson1972,Izhi}. This simple Additive Model can display different stationary solutions (one or three, depending on the system's parameters), but clearly the one dimensionality of the dynamical system forbids oscillatory dynamics. Interesting enough, a statistical treatment of the problem provides a picture in which the dynamics is two-dimensional \cite{Luke2013,Montbrio2015,Strogatz2016}. In this two dimensional description the asymptotic states are stationary points, and yet the transient can present oscillatory (damped) behavior. In this work, we explore the dynamical consequences of these frequencies present in the transients towards asymptotic stationary values of the ensemble's dynamics, subjecting the system to a variety of time dependent forcing. We will show that some observables describing the macroscopic outcomes of the system present resonant features.

We will explore the effect on the system's macroscopic features of three different forcing strategies. In the first case, forcing all units with the same strength. In the second case, the forcing strength on the units will follow a continuous distribution. In the last case under study, we will force a subpopulation of the original set of excitable units.

The work is organized as follows. The second section deals with the origin of the damping frequency in the autonomous system. The third section presents the derivation of the equations for the different forcing strategies. Finally, in the fourth section, we present the system's response to different forcing frequencies in the studied cases. Both by means of the analytic treatment and simulations of the problem, we find that, in the three cases, the system's global activity presents a maximum when the forcing frequency matches the ensemble's damping frequency (a value that depends on the distribution of parameters for units in the ensemble and their coupling weight).

\section{THE AUTONOMOUS MODEL}
\label{sec:2}

\subsection{Neurons}
\label{sec:2A}
Neurons are highly nonlinear systems whose principal characteristic is their excitability. They can be generally classified into two types depending on how the transition from rest to spiking state is made when an external current is applied \cite{Izhikevich2007}. Type-I neurons start firing at an arbitrarily low frequency rate whereas Type-II neurons start firing with finite frequency rate as soon as the threshold current is achieved. A canonical way for modeling a neuron is as an excitable oscillator, for example, the widely used Adler's equation \cite{Adler1946}:
\begin{equation}
\label{ThetaNeuron}
\dot{{\theta }}\left(t\right)=\frac{d{\theta }\left(t\right)}{dt}={w}-\mathrm{sin}\left({\theta }\right)+{I}_{in}
\end{equation}

\begin{figure}[!ht]
  \begin{center}
    \includegraphics[scale=.6]{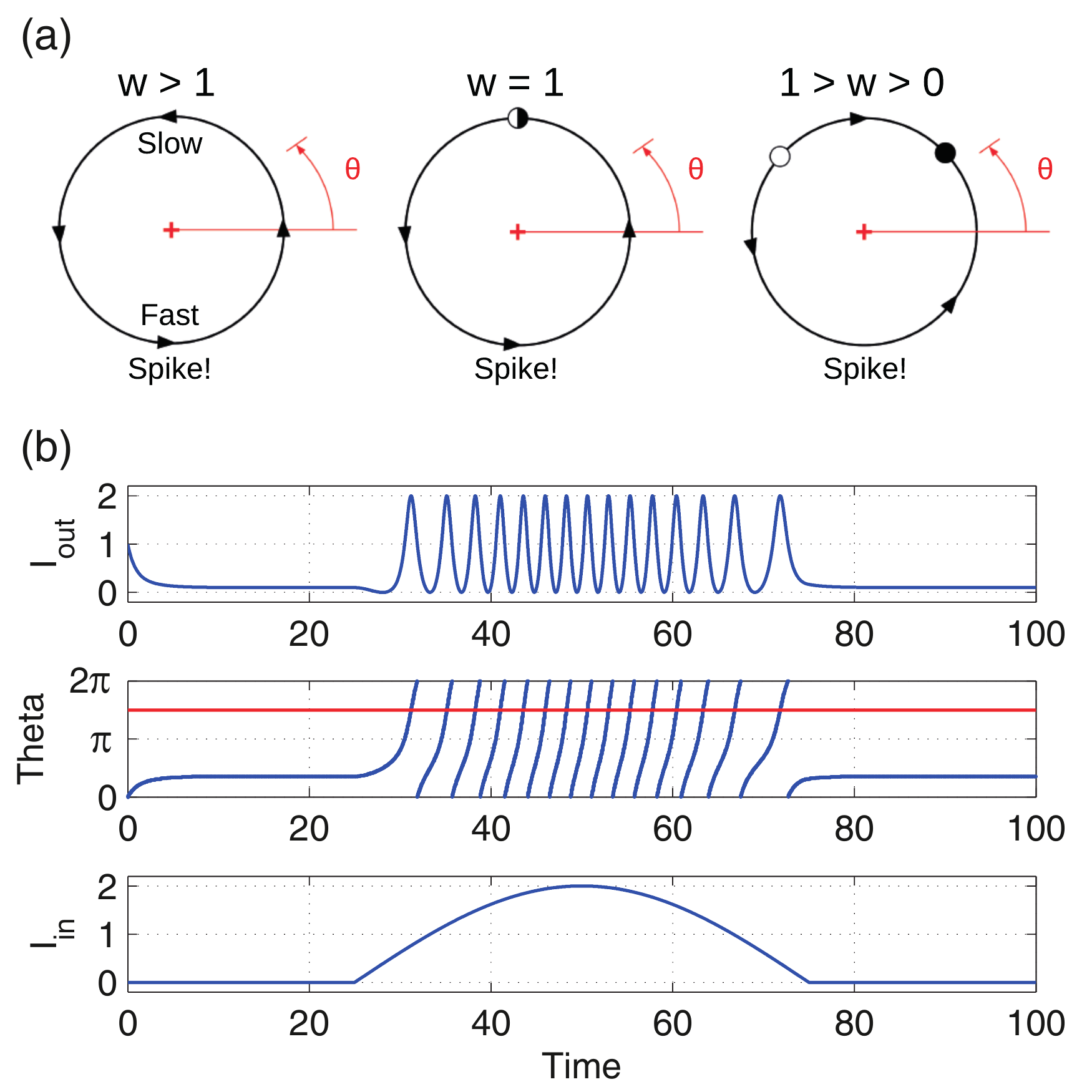} 
  \end{center}
  \caption{(a) Scheme of Saddle-Node bifurcation in Adler's equation. When $w$ is greater than 1, there are no fixed points, the oscillator is in spiking state. For $1>w>0$ a pair of stable and unstable fixed points are generated, so the oscillator will end in a resting state. (b) Adler neuron response to an external input current $I_{in}$. Note that when theta crosses the value $3\pi/2$, the output current ${I}_{out}=1-sin({\theta})$ reaches a maximum.}
 \label{fig_01}
\end{figure}

Despite being simple, this Adler neuron model reflects the capability of a neuron of being active or resting depending on an intrinsic parameter $w$ and, in case of connected, an injected current ${I}_{in}$. This equation presents a Saddle-Node in Limit Cycle bifurcation, so the oscillation frequency can be arbitrarily low as in a Type-1 neuron. Oscillator behavior is depicted schematically in Figure \ref{fig_01}a.

\subsection{Equations}
\label{sec:2B}
We consider for our analysis a population of N excitatory neurons, i. e. neurons that promote others to fire, coupled all-to-all with the same strength $k>0$. In this work, the coupling function between the neurons $i$ and $j$ depends only on the state of the presynaptic one and is given by the expression ${I}_{ij}=(1-sin({\theta }_{j}))$. The coupling function has it's maximum at $3\pi/2$, so we will consider that a neuron is spiking when crossing that value. This differs from the Kuramoto case, where coupling tends to bring closer units and depends on the relative position between them. The response of a single neuron to an external input current ${I}_{in}$ is presented in Figure \ref{fig_01}b, note that the generated current ${I}_{out}$ mimics a parabolic bursting.

The whole system of N equations has the form
\begin{equation}
\label{Velocidad}
\dot{{\theta }_{i}}\left(t\right)={w}_{i}-\mathrm{sin}\left({\theta }_{i}\right)+\sum _{j=1}^{N}\frac{k}{N}{I}_{ij}\left({\theta }_{j}\right)
\end{equation}
The intrinsic excitability parameter of the $i^{\text{\tiny th}}$ unit ($w_{i}$) is taken from a Lorentzian distribution which can be expressed as
\begin{equation}
\label{lore}
g\left(w \right)=\frac{\Delta}{2\pi\left(\left(w-{w}_{0}\right)^2+{\Delta}^2\right)},
\end{equation}
where parameter ${w}_{0}$ refers to the center of the distribution and $\Delta$ specifies the half-width at half-maximum. We consider only the physically relevant case ${w}_{0}>0$ and $\Delta>0$. Because of the distribution tail, there are still some non-physical units with negative $w$ values. Yet, the contribution of each of this units is $\sim \left(k/N\right){I}_{ij}\left({\theta }_{j}\right)$. Therefore, as long as ${w}_{0}>>\Delta$, the overall contribution of negative $w$ units will be neglectable.  

We will consider the limit $N \to \infty$. Within this regime, the state of the system can be described by means of a distribution function $f\left(w,\theta,t\right)$. This function is normalized to $1$ so that excitability parameter is distributed according to 
\begin{equation}
\label{CondDistrib}
\underset{0}{\overset{2\pi}{\int }}f\left(w,\theta,t\right)d\theta =g\left(w \right).
\end{equation}
Since the number of neurons in the system is conserved, the distribution function satisfies the continuity equation
\begin{equation}
\label{conti}
\frac{\partial f}{\partial t}+\frac{\partial }{\partial \theta }\left(\dot{\theta}f\right)=0.
\end{equation}
Using a mean-field approach \cite{Kuramoto1975,strogatz2000,dorfler2014}, it is possible to express the velocity of each oscillator as a function of the order parameter $r(t)$ by
\begin{equation}
\label{vel}
\dot{\theta}\left(w,\theta,t\right)=w-\mathrm{sin}\left(\theta\right)+k-kIm\left(r\left(t\right)\right),
\end{equation}
where $r(t)$ in the continuous limit is defined as
\begin{equation}
\label{order}
r\left(t\right)=\underset{-\infty}{\overset{\infty}{\int }}\underset{0}{\overset{2\pi}{\int }}f\left(w',\theta',t\right)e^{i\theta'}d\theta'dw'.
\end{equation}

Following the procedure proposed by Ott and Antonsen \cite{Ott2008}, we expand the distribution function in Fourier modes
\begin{equation}
\label{Distrib}
f\left(w,\theta,t\right) =\frac{g\left(w \right)}{2\pi}\left[1+\sum_{n=1}^{\infty}\left(f_{n}(w,t)e^{ni\theta}+f_{n}^{*}(w,t)e^{-ni\theta}\right)\right],
\end{equation}
and introduce the so-called OA anzat
\begin{equation}
\label{anzat} f_{n}(w,t)=\left[\alpha\left(w,t\right)\right]^n.
\end{equation}
It is remarkable that, if this assumption is made, the evolution of all the linearly independent modes $f_{n}(w,t)$ imposes the same condition on $\alpha\left(w,t\right)$
\begin{equation}
\label{alpha}
\dot{\alpha}\left(w,t\right)=-i\alpha\left(w,t\right)\left[w+k-kIm\left(r\left(t\right)\right)\right]-\frac{{\left(\alpha\left(w,t\right)\right)}^2}{2}+\frac{1}{2}.
\end{equation}

When integrating (\ref{order}) with respect to $\theta$, all $e^{mi\theta}$ terms with $m\neq0$ will vanish.
Following \cite{Ott2008}, we require that $\alpha(w,t)$ can be analytically continued from real $\omega$ into the complex $\omega$-plane, that this continuation has no singularities in the lower half $\omega$-plane, and that $\alpha\left(w,0\right) \to 0$ when ${w}_{I}\to -\infty$. If these conditions are satisfied , then it is possible to solve the $\omega$ improper integral via the Residue Theorem, obtaining
\begin{equation}
\label{order2}
r\left(t\right)=\underset{-\infty}{\overset{\infty}{\int }}\alpha^{*}\left(w,t\right)g(w)dw=\alpha^{*}\left(w_{0}+i\Delta,t\right).
\end{equation}
In this way, the time evolution of the order parameter $r\left(t\right)$ will be governed by the expression obtained when replacing $w=w_{0}+i\Delta$ in the complex conjugate of (\ref{alpha}), namely,
\begin{equation}
\label{dinamica2}
\dot{r}\left(t\right)=ir\left(t\right)\left[w_{0}+i\Delta+k-kIm\left(r\left(t\right)\right)\right]-\frac{{\left(r\left(t\right)\right)}^2}{2}+\frac{1}{2}.
\end{equation}

Notice that, on the continuous limit and under the OA anzat, it is not necessary to solve the whole system of N coupled differential equations in order to know the value of the mean-field quantity $r\left(t\right)$. The dimension of the problem drastically collapses to a one dimensional nonlinear complex differential equation.

\subsection{Measure of the activity}
\label{sec:2C}

The level of \emph{activity} of a neural population is a pertinent way of describing it, particularly if the population is in charge of a macroscopic motor action. Following previous work \cite{Roulet2016}, we define the activity as the Mean Firing Rate $X\left(t\right)$, i.e. the average number of spikes in the population per unit of time. Considering that our neurons fire at $\theta=\frac{3}{2}\pi$, this can be expressed as
\begin{equation}
\label{disparo}
X\left(t\right)=\underset{-\infty }{\overset{\infty }{\int }}{\left(f\left(w,\theta,t\right)\stackrel{.}{\theta}\left(w,\theta,t\right)\right)|}_{\theta=\frac{3}{2}\pi}dw.
\end{equation}
Assuming the same conditions previously imposed on $\alpha\left(w,A,t\right)$, and following the procedure presented in \cite{Roulet2016}, it is possible to solve this integral (the procedure is detailed in Section \ref{sec:3}) obtaining

\begin{equation}
\label{Xt_1}
X(t)=\frac{1}{2\pi} \left(  w_{0}+1+k-kIm\left[r\left(t\right)\right]\right)+
2 Re \Bigg[\frac{1}{2\pi}\left(\frac{r^{*}(t)}{i-r^{*}(t)}\right)\bigg(w_{0}-i \Delta
+1+k-kIm \left[r\left(t\right)\right] \bigg) \Bigg].
\end{equation}

\subsection{Bifurcation Diagram}
\label{sec:2D}
\begin{figure}[!ht]
  \begin{center}
    \includegraphics[width=\textwidth]{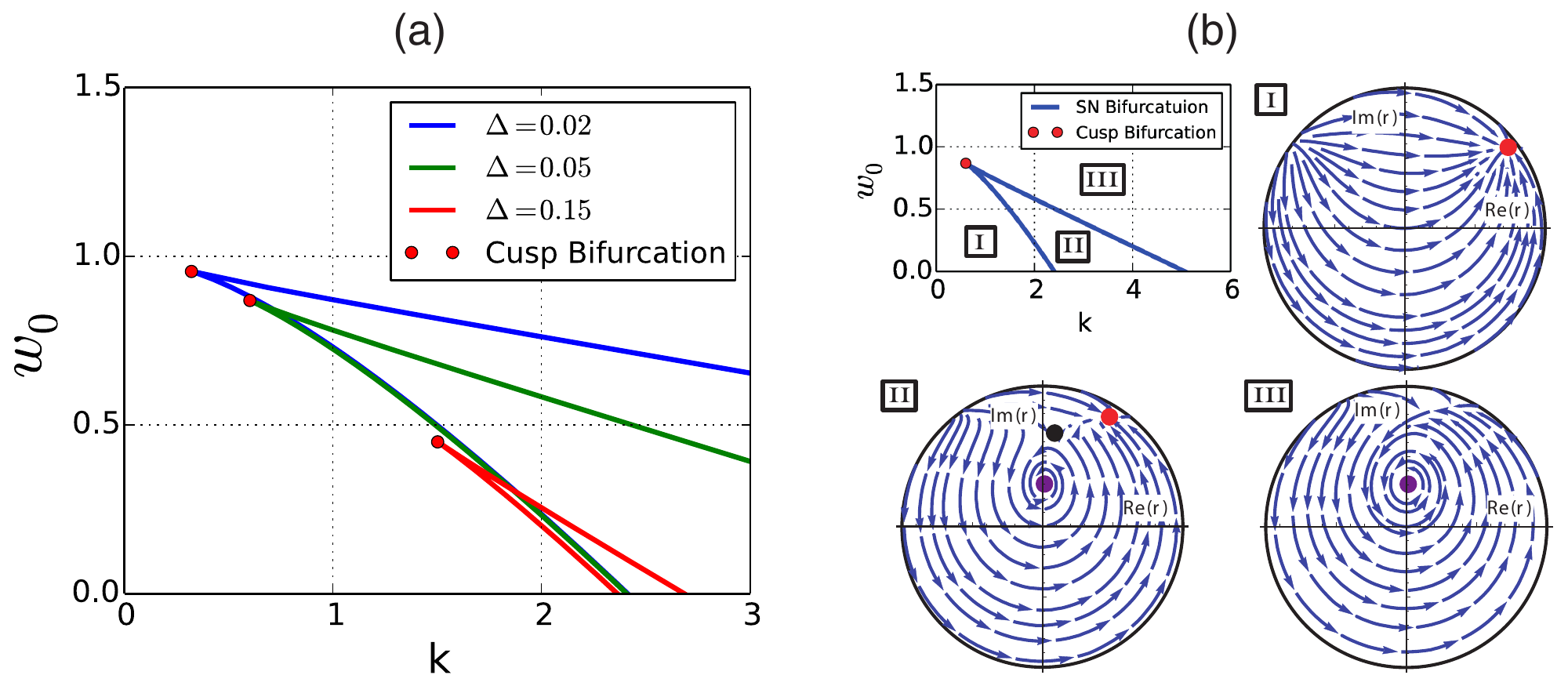} 
  \end{center}
  \caption{Bifurcation analysis of the autonomous model. (a) Bifurcation diagram for the system in the $(k,w_{0})$ plane for different values of $\Delta$. Curves denote Saddle-Node bifurcations and red dots denote Cusp bifurcations. (b) Phase portraits for different zones of the parameter space in the case $\Delta=0.05$: Region I with $(k,w,\Delta) = (0.5,0.4,0.05)$, Region II with $(k,w,\Delta) = (2,0.4,0.05)$ and Region III with $(k,w,\Delta) = (1.5,0.75,0.05)$. Fixed points are represented by dots with color red for stable non-rotational, black for Saddle Node and violet for stable rotational.}
 \label{fig_02}
\end{figure}

The analysis of the bifurcation diagram for the autonomous model was performed using \emph{PyDSTool} Python toolbox \cite{PyDSTool} \cite{PyDSTool2}. As stated before, the dynamics of $r\left(t\right)$ is governed by the complex equation (\ref{dinamica2}), which is equivalent to a system of two real non-linear coupled differential equations. The bifurcation diagram for the parameter region of interest, $\Omega=(k>0,w_{0}>0,\Delta>0)$, is shown in Figure \ref{fig_02}a. Within this region, the only limit sets presented by the system are fixed points. In terms of $(k,w_{0})$ planes, if $\Delta<\Delta_c\approx0.226$, a cusp bifurcation takes place and two Saddle-Node curves delimit a region with 3 fixed points. The absence of limit cycles in $\Omega$ is not surprising since, despite being two-dimensional, there are no competing forces in the system: all couplings are impulsive and almost all units have positive intrinsic frequencies. The bifurcation diagram for this autonomous model results equivalent to that found for similar models in previous studies \cite{Luke2013,Montbrio2015,Strogatz2016}.

As shown in Figure \ref{fig_02}b, the zone delimited by the bifurcation curves is bistable: it presents one stable spiral and one stable node connected by a slow manifold and separated by a saddle point. Each Saddle-Node bifurcation curve corresponds to the Saddle point collapsing against one of the stable solutions, leaving the other one as the only attractor in the system. When compared in terms of their activities, the stable spiral is the high activity solution whereas the stable node is the low activity solution.

As mentioned in the introduction, one of the simplest empirical models for activity in population of excitatory neurons is the Additive Model:
\begin{equation}
\label{Adit}
\dot{a}\left(t\right)=-a+S(\rho + c a),
\end{equation}
where $S$ is the sigmoid function $S(y)=1/(1+e^{-y})$, $a\left(t\right)$ is the activity of the population, $c$ is a feedback parameter and $\rho$ accounts for an overall external input. It is a Wilson-Cowan model for only one excitatory population \cite{Wilson1972,Izhi}.

\begin{figure}[!ht]
  \begin{center}
    \includegraphics[scale=.6]{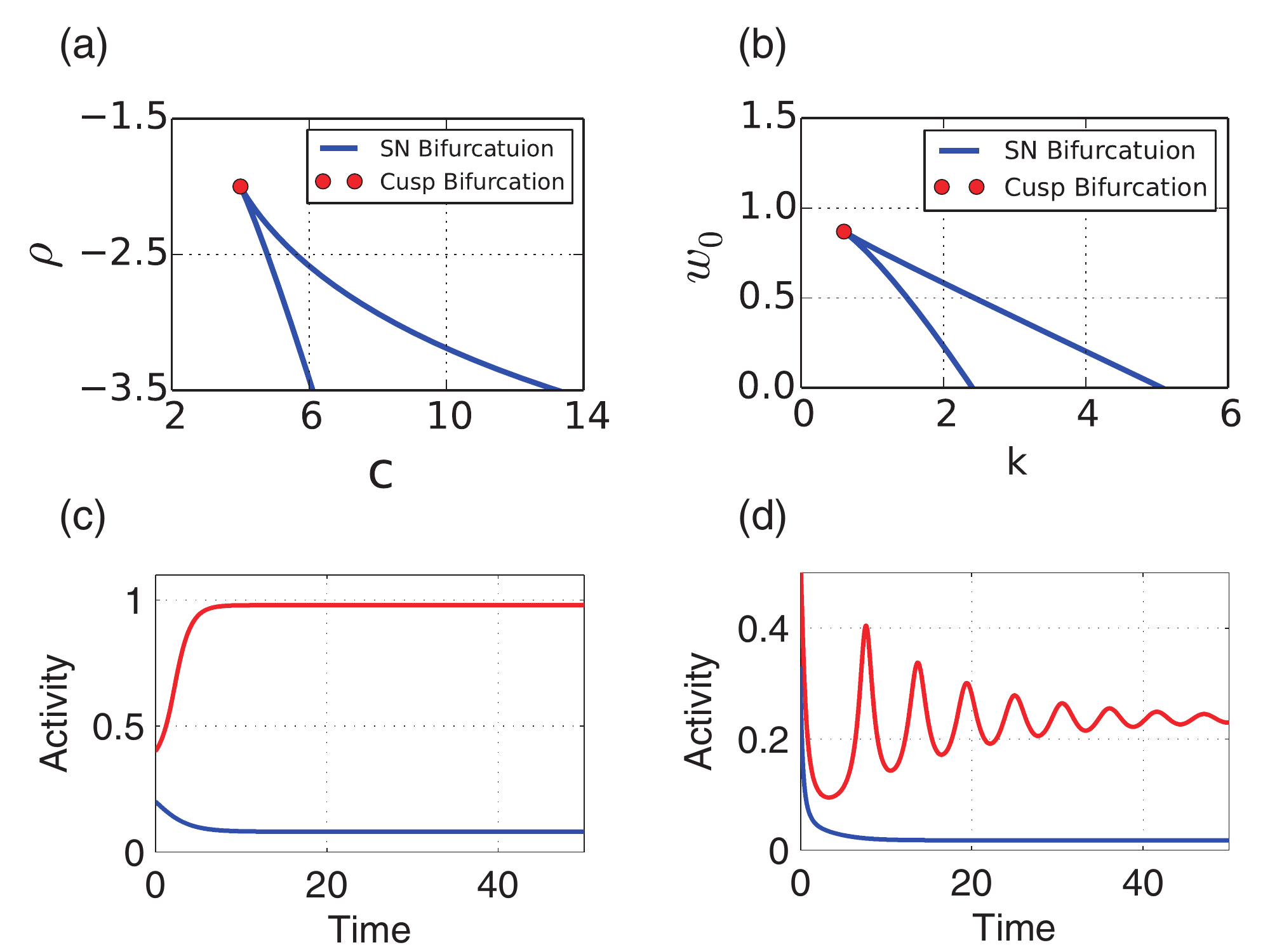} 
  \end{center}
  \caption{(a) Additive model bifurcation diagram. (b) Autonomous model's bifurcation diagram for $\Delta = 0.05$. (c) Low (Blue) and high (Red) activity solutions in the bistable region of the Additive Model. (d) Low (Blue) and high (Red)  activity solutions in the bistable region of the autonomous model.}
 \label{fig_03}
\end{figure}

The bifurcation diagram of the Additive model is shown in Figure \ref{fig_03}. Compared with the model constructed from first principles, if the Lorentz distribution is narrow enough, both present the same type of bifurcations: Two Saddle-Node curves that collide in a Cusp Bifurcation. Also, both models have a high and a low activity stable solution in the bistable region.

The main difference between both activity models is that one is governed by a 2-dimensional system and can present a transitory underdamped oscillatory solution, while oscillations are impossible for the one-dimensional Additive model. In section \ref{sec:4}, we will show that the frequency of this rotational solution is of great interest when the system is connected to an external stimulation.

\subsection{Rotational solution}
\label{sec:2E}

\begin{figure}[!ht]
  \begin{center}
    \includegraphics[width=\textwidth]{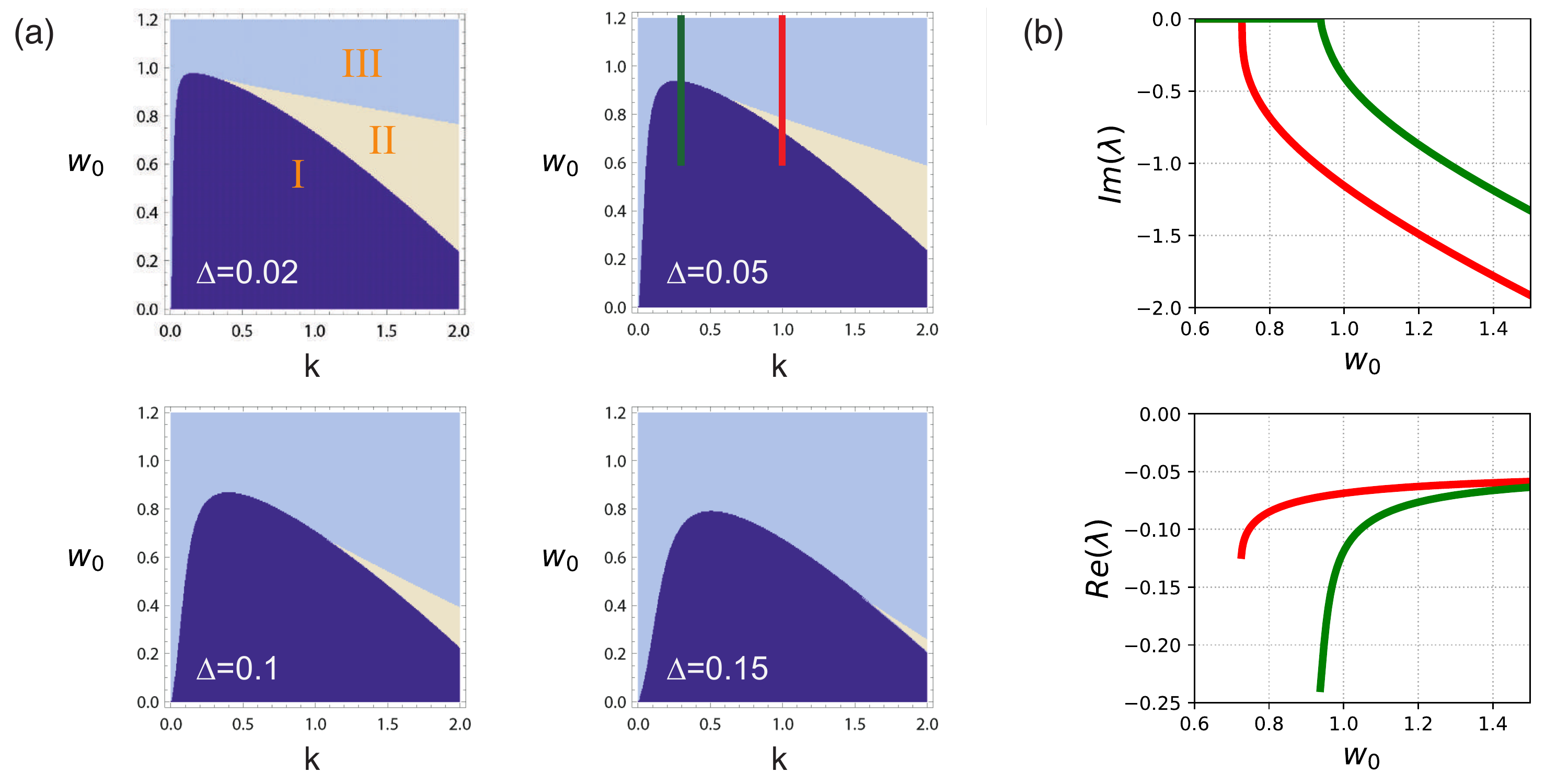} 
  \end{center}
  \caption{(a) Types of solutions in the $(k,w_0)$ plane for different $\Delta$ values in the autonomous model. Region I (purple) corresponds to the stable node solution, Region II (beige) to the bistable zone and Region III (light blue) to the stable spiral solution. (b) Imaginary and real parts of the complex eigenvalue for the linearized system about the rotational solution in autonomous model as a function of $w_0$. Curves showed are for $\Delta = 0.05$ and two different $k$ values: $k=0.3$ (green curve) rotational solution born in a transition from region $I$ to $III$, and $k=1$ (red curve) born in transition from $I$ to $II$. Both curves are also depicted in (a).}
 \label{fig_04}
\end{figure}

In order to delimitate the parameter region where the system presents a rotational solution, regions $II$ and $III$ of Figure \ref{fig_02}b, we computed the eigenvalues of the linearized system around the fixed points and checked where the imaginary part was non-zero. Results in terms of $(k,w_0)$ planes for different values of $\Delta$ are presented in figure \ref{fig_04}a.

Note that if the system is uncoupled, i.e. $k=0$, it always presents a unique rotational solution. This leads us to the important observation that the rotational behavior can be achieved by the heterogeneity in the oscillators population on its own, characterized by the Lorentzian distribution, even in the absence of coupling. In terms of the full system, this transient before reaching the fixed point corresponds to the evolution of the distribution function $f\left(w,\theta,t\right)$ while approaching asymptotically its stable form $\tilde{f}\left(w,\theta\right)$. 

It is remarkable that, even in the uncoupled case, the oscillators tend to fire simultaneously generating activity's maxima and minima. This coordination is due to the oscillators getting stuck together in the saddle bifurcation ghost near $\theta=\pi/2$. For oscillators with $w$ slightly larger than 1, the time they spend passing through that zone is much longer than the one they spend in exploring the rest of the circle. As time evolves, the differences in their intrinsic frequencies $w$ set them apart and the activity's oscillation fades away. An arbitrary long time after the initial conditions, the probability of finding an oscillator with intrinsic frequency $w$ in some region $d\theta$ of the circle is inversely proportional to its velocity $v(w,\theta)=w-\sin(\theta)$ in that region. 

\begin{figure}[!ht]
  \begin{center}
    \includegraphics[width=\textwidth]{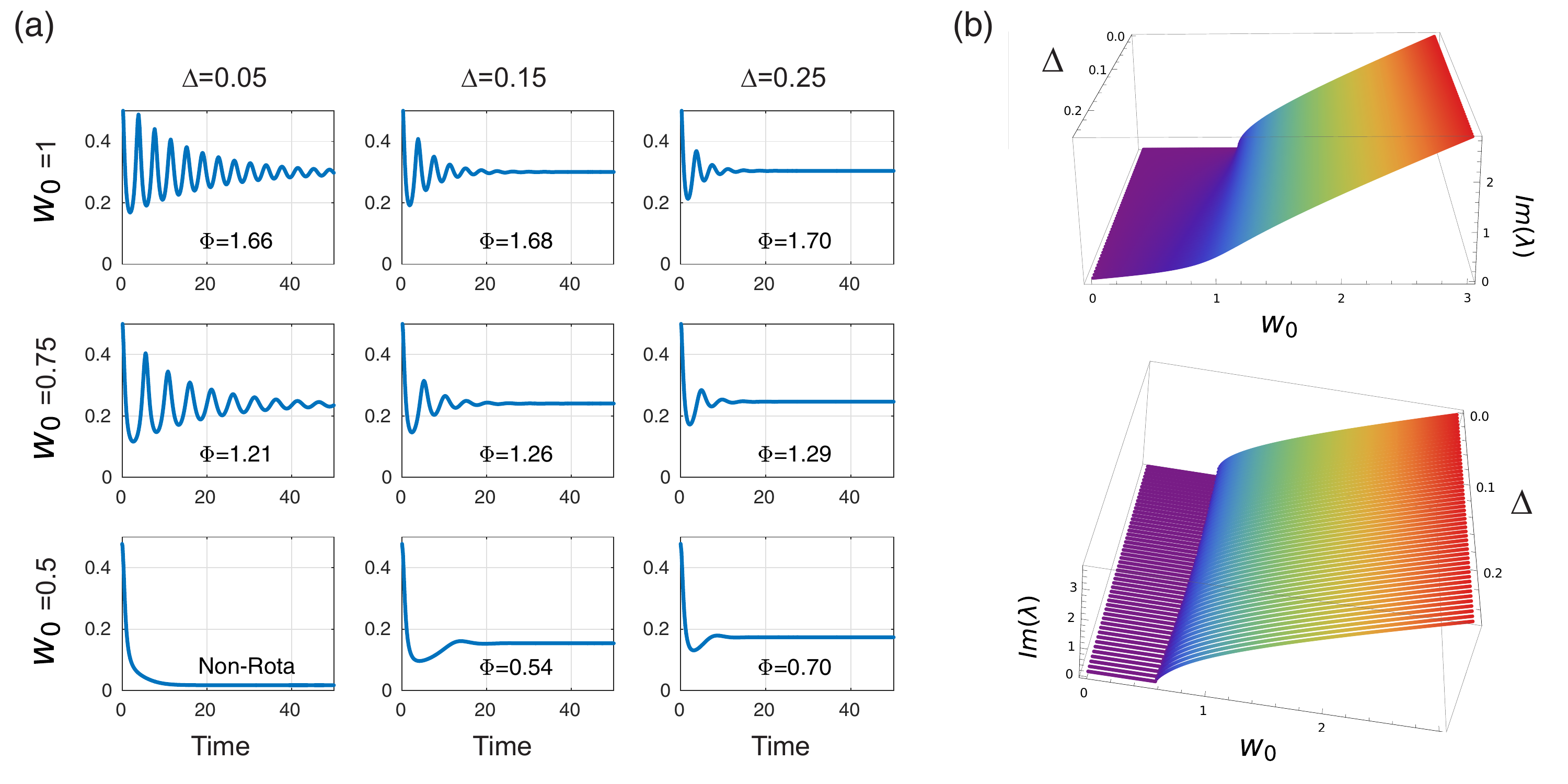} 
  \end{center}
  \caption{(a) Activity $X(t)$ of the autonomous model when approaching rotational fixed point (if it exists) for different $w_0$ and $\Delta$ values in the $k=1.5$ plane. The angular frequency ($\Phi$) is reported for each of the cases. (b) Modulus of the imaginary part of $\lambda$ as a function of $w_0$ and $\Delta$ in the uncoupled case $k=0$ (up) and the coupled case $k=1$ (down). $\lambda$ stands for the complex eigenvalue of the linearized system about the rotational solution in the autonomous model.}
 \label{fig_05}
\end{figure}

Figure \ref{fig_04}b shows the value of the imaginary and real parts of complex eigenvalue $\lambda$ as a function of $w_0$ for two different trajectories. $\lambda$ is the eigenvalue of the linearized system about the rotational fixed point. In one trajectory, the system crosses from region $I$ to region $II$ going through a saddle node bifurcation and, in the other, it goes from region $I$ to $II$ without crossing any bifurcation. In both cases this eigenvalue is born with $Im(\lambda)=0$, meaning that the dynamic near this fixed point, in the proximity of region's $I$ frontier, has an arbitrarily low frequency associated to it. Also, in that conditions, the real part of the eigenvalue $Re(\lambda)$ presents its larger absolute value. So, in that regime, solutions will be strongly attracted to the fixed point, generating smaller amplitude oscillations. In Figure \ref{fig_05}a we present the activity of the system falling to the rotational fixed point for different values of distribution parameters $w_0$ and $\Delta$, when $K=1.5$. It shows that, for smaller $w_0$ values, the damped oscillation frequency decreases. Also, as stated before, larger differences in units frequencies (larger $\Delta$ value) increase the damping rate of oscillations.

In conclusion, the relation between the angular frequency of the rotational solution, $\Phi$, and the parameters of the system, $(w_0,k,\Delta)$, can be computed via the value of $Im(\lambda)$. Figure \ref{fig_05}b exhibits this relation for two different $k$ values. It is important to stress that the value of $Im(\lambda)$ varies not only with $w_0$ and $k$, but also with $\Delta$. Even in the uncoupled case $k=0$, its value is not completely determined by the value of $w_0$, particularly in the region where $w_0 \approx 1$. This is because, near this value, the distribution width is decisive in determining how many oscillators have $w \in [-1,1]$, region of its quiescent state. So, this damped oscillation frequency is a population property that depends on the coupling strength and the form of the $w$ distribution function, not only the location of it's maximum $w_0$.

\section{THE FORCED MODEL}
\label{sec:3}

\subsection{Equations}

We will now consider an external input current acting on the system. This input may represent the activity of other neural population whose neurons project towards ours. In a real nervous system it is unlikely to have all the neurons of some neural nucleus homogeneously connected to another. There will be neurons more influenced by the activity of that external nucleus than others.
To address this possibility, we propose a model in which the velocity of each unit is given by
\begin{equation}
\label{VelForzadoHete}
\dot{{\theta }_{i}}\left(t\right)={\omega }_{i}-\mathrm{sin}\left({\theta }_{i}\right)+\sum _{j=1}^{N}\frac{k}{N} {I}_{ij}\left({\theta }_{j}\right)+{A}_{i} sin\left(\Omega t\right),
\end{equation}
where $\Omega$ is the frequency of the forcing term and ${A}_{i}$ its amplitude, which is taken from a distribution $\Gamma(A)$. We will analyze three different cases and we will show that, the mean field  of these systems is captured by a reduced set of coupled differential equations.

\subsubsection{Homogeneously forced}
\label{sec:3A}

As a first case we will consider $\Gamma_1=\delta(A-A_0)$, meaning that all units are being homogeneously forced with the same strength $A_0$. In this case, the procedure to follow is completely analogue to the one presented in section \ref{sec:2B}. The dynamics of the order parameter will be governed by the equation

\begin{equation}
\label{dinamica_heter_1}
\dot{r}\left(t\right)=  ir\left(t\right)\Big[ w_{0}+i\Delta+k-kIm\left(r\left(t\right)\right)+{A}_{0}\ sin\left(\Omega t\right)\Big]-\frac{{\left(r\left(t\right)\right)}^2}{2}+\frac{1}{2}.
\end{equation}

\subsubsection{Heterogeneously forced with continuous distribution}
\label{sec:3B}

As a second forcing strategy, we will consider that $\Gamma(A)$ follows a lorentzian distribution
\begin{equation}
\label{lore_A}
\Gamma_2\left(A \right)=\frac{\Delta_A}{2\pi\left(\left(A-{A}_{0}\right)^2+{\Delta_A}^2\right)},
\end{equation}
where $A_0$ is the center of the distribution and $\Delta_A$ its half width at half maximum.

In this new system, the distribution function also depends on the value of $A$ and satisfies 
\begin{equation}
\label{CondDistribHeter}
\underset{0}{\overset{2\pi}{\int }}f\left(w,A,\theta,t\right)d\theta =g\left(\omega \right)\Gamma_2(A),
\end{equation}
as well as the continuity equation (\ref{conti}). Making a Fourier decomposition of the distribution function and restricting it to the OA anzat (\ref{anzat}), it can be expressed as
\begin{equation}
\label{DistribHeter2}
f\left(w,A,\theta,t\right)=\frac{g\left(\omega \right) \Gamma_2(A)}{2\pi}\Bigg[1+\sum_{n=1}^{\infty}\Big(\alpha(w,A,t))^n e^{ni\theta}+(\alpha^{*}(w,A,t))^n e^{-ni\theta}\Big)\Bigg].
\end{equation}
The evolution of each Fourier mode, given by the continuity equation (\ref{conti}) and the velocity (\ref{VelForzadoHete}), result in a unique condition
\begin{equation}
\label{alphaForzadoHeter}
\dot{\alpha^*}\left(w,A,t\right)=i\alpha^*\left(w,A,t\right)\big[w+k-kIm\left(r\left(t\right)\right)+A\ sin\left(\Omega t\right) \big]-\frac{{\left(\alpha^*\left(w,A,t\right)\right)}^2}{2}+\frac{1}{2}.
\end{equation}

Also, the order parameter $r(t)$ can be written as
\begin{equation}
\label{OrderHeter}
r\left(t\right)=\underset{-\infty}{\overset{\infty}{\int }}\underset{-\infty}{\overset{\infty}{\int }}\underset{0}{\overset{2\pi}{\int }}f\left(w',A',\theta',t\right)e^{i\theta'}d\theta'dw'dA'
=\underset{-\infty}{\overset{\infty}{\int }}\underset{-\infty}{\overset{\infty}{\int }}\alpha^*\left(w',A',t\right)g(w')\Gamma_2(A')dw'dA'.
\end{equation}
Imposing on $\alpha$ the same requirements for the variable $A$ as the ones imposed for $\omega$ in the previous section, it is possible to solve both integrals, obtaining 
\begin{equation}
\label{OrderHeterLargo}
r \left(t\right) = \alpha^*\left({w}_{0}+i\Delta,{A}_{0}+i\Delta_A,t\right).
\end{equation}
Replacing on (\ref{alphaForzadoHeter}), the time evolution of the order parameter results 
\begin{equation}
\label{dinamica_heter_1}
\dot{r}\left(t\right)=  ir\left(t\right)\Big[w_{0}+i\Delta+k-kIm\left(r\left(t\right)\right)+\left({A}_{0}+i\Delta_A\right)\ sin\left(\Omega t\right)\Big]-\frac{{\left(r\left(t\right)\right)}^2}{2}+\frac{1}{2}.
\end{equation}
So, in this framework, the heterogeneous input to the units of the system can be thought of as a temporal variation of the $\omega$ and $\Delta$ parameters.

This procedure can be used for any $A$ distribution, as long as the integral (\ref{OrderHeter}) is analitically solvable. 

\subsubsection{Heterogeneously forced with discrete distribution}
\label{sec:3C}

As the last case, we will consider a distribution $\Gamma_3(A)=(1-p) \ \delta(A)\ +\ p \ \delta(A-{A}_{0})$ with $0\leq p\leq1$ . Namely, neurons has a probability $p$ to be forced by a external input with amplitude ${A}_{0}$, and probability $1-p$ of not being influenced by it. in this case, the order parameter will be given by the expression
\begin{equation}
\label{OrderHeterLargo}
\begin{split}
r \left(t\right)=\underset{-\infty}{\overset{\infty}{\int }}\alpha^*\left({w}_{0}+i\Delta,A',t\right)\Gamma_3(A')dA'
&=(1-p)\ \alpha^*\left({w}_{0}+i\Delta,0,t\right) + p\ \alpha^*\left({w}_{0}+i\Delta,{A}_{0},t\right) \\
&=(1-p)\ y(t) \  + \ p\ z(t).
\end{split}
\end{equation}
In this last expression, the order parameters of the population of disconnected and connected units are identified as $y(t)$ and $z(t)$ respectively. The evolution of this sub-order parameters can be obtained from (\ref{alphaForzadoHeter}). Because of the $Im(r(t))$ term, their dynamic is coupled. So, in order to compute $r(t)$, the following system of two ordinary complex nonlinear differential equation must be solved
\begin{equation}
\label{EcusHeter}
\begin{split}
\dot{y}\left(t\right)&=iy\left(t\right) \big\{{w}_{0}+i\Delta+k-kIm\left[(1-p)\ y(t)+p\ z(t)\right] \big\} -\frac{\left(y\left(t\right)\right)^2}{2}+\frac{1}{2},
\\
\dot{z}\left(t\right)&=iz\left(t\right) \big\{{w}_{0}+i\Delta+k-kIm\left[(1-p)\ y(t)+p\ z(t)\right] \big\}+ {A}_{0}\ sen\left(\Omega t\right)-\frac{\left(z\left(t\right)\right)^2}{2}+\frac{1}{2}.
\end{split}
\end{equation}
Notice that, if $p=0$, there are no units connected to the forcing input, and the dynamic of the system is reduced to one complex differential equation, equivalent to (\ref{dinamica2}). On the other side, if $p=1$, we have all units connected to the external current and, again, the dynamic can be obtained through only a one-dimensional complex differential equation.

An important observation is that the number of resulting equations, i.e. the dimensionality of the macroscopic dynamics, is not directly related to the level of heterogeneity of $\Gamma(A)$. For instance, the distribution $\Gamma_2(A)$ allows more possible values for $A$ than $\Gamma_3(A)$ and yet the system to be solved has a smaller number of equations. For continuous distributions where the integral (\ref{OrderHeter}) is solvable, the number of resulting equations is equal to the number of poles in the lower half plane of $\Gamma(A)$. But for discrete distributions, the number of equations does increase with the number of possible $A$ values, and so does the number of identifiable populations in the system. E.g., for $\Gamma_3(A)=(1-p-q)\ \delta(A-{A}_{1})+p\ \delta(A-{A}_{2})+q\ \delta(A-{A}_{2})$ with $0\leq p \leq 1$, $0\leq q \leq 1$ and $0\leq p+q \leq 1$, the system to be solved is composed by three coupled complex differential equations, one for each interacting population.

\subsection{Mean Firing Rate calculation}
\label{sec:3D}

\begin{figure}[!ht]
  \begin{center}
    \includegraphics[scale=.7]{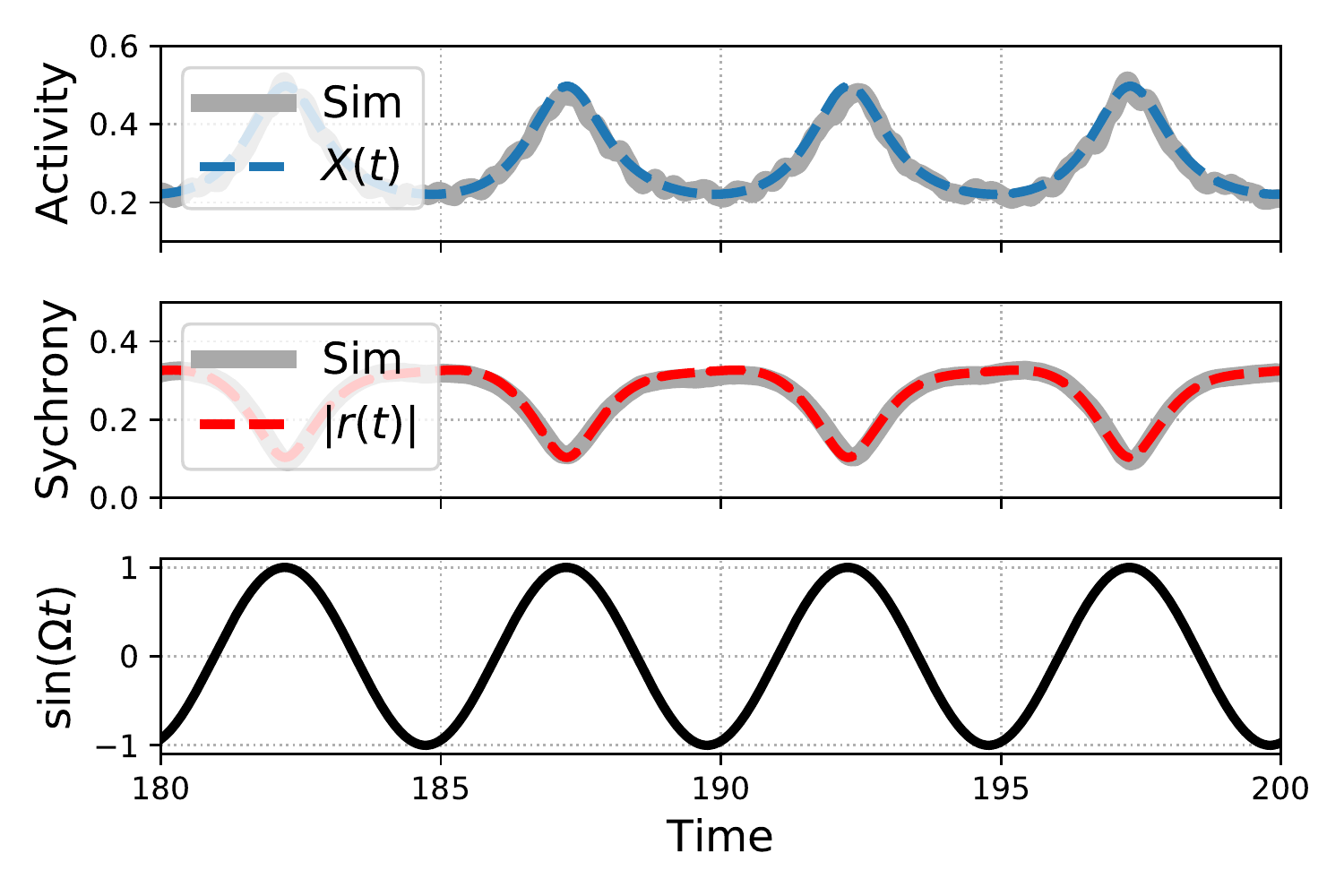} 
  \end{center}
  \caption{Comparison between the analytical value of the Mean Firing Rate $X\left(t\right)$ and the synchronicity $|r\left(t\right)|$ with those same quantities calculated from the simulation of the whole system. Simulation was performed with N=12000 oscillators and an integration time step of 0.001 for an Heterogeneously forced system with parameters $(k,w_{0},\Delta,\Omega,A_{0},\Delta_A)=(1.5,1.1,0.05,1.25,0.4,0.025)$.}
 \label{fig_06}
\end{figure}

The level of activity in these forced models will be given by the Mean Fire Rate
 
\begin{equation}
\label{disparo_heter}
X\left(t\right)=\underset{-\infty }{\overset{\infty }{\int }}{\left(f\left(w,\theta,A,t\right)\stackrel{.}{\theta}\left(w,\theta,A,t\right)\right)|}_{\theta=\frac{3}{2}\pi}dw.
\end{equation}
Assuming the same conditions previously imposed on $\alpha\left(w,A,t\right)$ it is again possible to solve this integral following the procedure presented in \cite{Roulet2016}.

Evaluating the distribution function (\ref{Distrib}) in $\theta = 3\pi/2$ it results
\begin{equation}
f\left(w,A,\frac{3}{2}\pi,t\right) =\frac{g\left(w \right)\Gamma (A)}{2\pi}\Bigg[1+\sum_{n=1}^{\infty}\big[\left(-i\alpha(w,A,t)\right)^n+\left(i\alpha^{*}(w,A,t)\right)^n\big]\Bigg].
\end{equation}
It's necessary for $\alpha(w,A,t)$ to fulfil $|\alpha(w,A,t)|<1$, if not, distribution $f\left(w,A,\theta,t\right)$ would diverge. Then, the summation can be reduced using the geometric series identity
\begin{equation}
\sum_{n=1}^{\infty}x^{n}=\frac{x}{1-x},
\end{equation}
resulting in
\begin{equation}
\label{Distrib4}
f\left(w,A,\frac{3}{2}\pi,t\right) =\frac{g\left(w \right)\Gamma (A)}{2\pi}\Bigg[1+\frac{\alpha\left(w,A,t\right)}{i-\alpha\left(w,A,t\right)}+\frac{\alpha^{*}\left(w,A,t\right)}{-i-\alpha^{*}\left(w,A,t\right)}\Bigg].
\end{equation}

Velocity can be also evaluated in $\theta = 3\pi/2$, obtaining 
\begin{equation}
\stackrel{.}{\theta}\left(w,A,\frac{3}{2}\pi,t\right)=w+1+k-kIm\left[r\left(t\right)\right]+ A \ sin(\Omega t)=w+ A \ sin(\Omega t) +Q(t),
\end{equation}
where the part of the velocity that doesn't depend on $w$ or $A$ is called $Q(t)$. 

Taking into account the above-mentioned, the activity $X(t)$ can be expressed as
\begin{equation}
\label{disparo2}
X\left(t\right)=\underset{-\infty }{\overset{\infty }{\int }}\Gamma\left(A \right) \Bigg\{\underset{-\infty }{\overset{\infty }{\int }}\frac{g\left(w \right)}{2\pi} \Bigg[ 1+\frac{\alpha\left(w,t\right)}{i-\alpha\left(w,t\right)} +\frac{\alpha^{*}\left(w,t\right)}{-i-\alpha^{*}\left(w,t\right)}\Bigg]\left(w+Q\right)dw \Bigg\}dA.
\end{equation}
The improper integrals over the variable $w$, the one between curly brackets, can be solved using the Residue theorem (see \cite{Ott2008} for details of the procedure to follow and \cite{Roulet2016} for comments on the divergence). Calling $S_{1}$, $S_{2}$ and $S_{3}$ to the three terms between non-curly brackets, the resulting expressions are:
\begin{equation}
\label{S123}
\begin{split}
S_{1}=&\frac{w_{0}+Q}{2\pi}, \\
S_{2}=&\frac{1}{2\pi}\left(\frac{\alpha\left(w_{0}-i\Delta,A,t\right)}{i-\alpha\left(w_{0}-i\Delta,A,t\right)} \right) \Big( w_{0}-i\Delta+A\ sin(\Omega t) +Q \Big),\\
S_{3}=&\frac{1}{2\pi}\left(\frac{\alpha^{*}\left(w_{0}+i\Delta,A,t\right)}{-i-\alpha^{*}\left(w_{0}+i\Delta,A,t\right)}\right) \Big( w_{0}+i\Delta+A \ sin(\Omega t) +Q \Big).
\end{split}
\end{equation}
These terms have to be integrated over the forcing strength variable $A$, obtaining a different result for each forcing strategy. 

\subsubsection{Homogeneously forced}

In this first case, $\Gamma_1 = \delta(A-A_0)$, the integration over the variable $A$ results trivial. Then, the activity can be obtained by simply summing the three terms and using the relation (\ref{OrderHeter}) between $\alpha$ and $r$, obtaining

\begin{equation}
\begin{split}
\label{XtForzado}
X(t)=&\frac{1}{2\pi} \left[w_{0}+1+k-kIm\left[r\left(t\right)\right]+A\left(1-sen\left(\Omega t\right)\right) \right]\\
&+2 Re\bigg[\frac{1}{2\pi}\left(\frac{r^{*}(t)}{i-r^{*}(t)}\right)\Big(w_{0}-i\Delta+1+k-kIm\left[r\left(t\right)\right]+A\left(1-sen\left(\Omega t\right)\right)\Big)\bigg].
\end{split}
\end{equation}

Note that $S1_{t}$ is real and $S_{3}={S_{2}}^{*}$, so $X(t)$ results real, as expected from it's definition. This will be the case for the three distributions.

\subsubsection{Heterogeneously forced with continuous distribution}

In the second case, $\Gamma_2(A)$ follows a Lorentzian distribution. In order to the solve the integral, the Residue Theorem can be used again obtaining

\begin{equation}
\label{Xt2}
\begin{split}
X_2(t)=&\frac{1}{2\pi} \left(w_{0}+A_0 \sin(\Omega t) +1+k-kIm\left[r\left(t\right)\right]\right)+ \\
& 2 Re \Bigg[ \frac{1}{2\pi}\left(\frac{r^{*}(t)}{i-r^{*}(t)}\right)\bigg( w_{0}+A_0 \sin(\Omega t)-i \left( \Delta+\Delta_A \  \sin(\Omega t)\right) +1+k-kIm\left[r\left(t\right)\right]\bigg) \Bigg].
\end{split}
\end{equation}

\subsubsection{Heterogeneously forced with discrete distribution}

In this last case, the activity of the whole system can be expressed as the sum of the activity of both subpopulations
\begin{equation}
\label{XtHetero}
X_3(t)=(1-p) \ X_y(t)  + p \ X_z(t).
\end{equation}
where
\begin{equation}
\begin{split}
\label{DefI2}
X_y(t)=&\frac{1}{2\pi} \left[w_{0}+1+k-kIm\left[r\left(t\right)\right]\right]
+ 2Re\Bigg[\frac{1}{2\pi}\left(\frac{y^*(t)}{i-y^*(t)}\right)\Big(w_{0}-i\Delta+1+k
-kIm\left[r\left(t\right)\right]\Big) \Bigg],\\
X_z(t)=&\frac{1}{2\pi}\left[w_{0}+1+k-kIm\left[r\left(t\right)\right]+A_{0}\left(1-sen\left(\Omega t\right)\right)\right]\\
&+2Re\Bigg[\frac{1}{2\pi}\left(\frac{z^*(t)}{i-z^*(t)}\right)\Big(w_{0}-i\Delta+1
+k-kIm\left[r\left(t\right)\right]+A_{0}\left(1-sen\left(\Omega t \right) \right) \Big) \Bigg].
\end{split}
\end{equation}

In figure \ref{fig_06} we present the time evolution of $X\left(t\right)$ and $|r\left(t\right)|$ for a population of oscillators forced with $\Gamma_2(A)$. Results computed from the reduced equations (\ref{dinamica2}) via the expressions (\ref{Xt2}) agree with those obtained through computational simulation of the whole system.

\section{RESULTS}
\label{sec:4}

In this section, we will explore the response of the system to the different forcing strategies. When forcing the system, in any of the three studied cases, two different limit sets are possible for the order parameter $r(t)$: limit cycles and chaotic attractors. We will focus our analysis to regions of parameters where the forcing does not elicit chaotic behavior.

\subsubsection{Homogeneously forced}

\begin{figure}[!ht]
  \begin{center}
    \includegraphics[width=\textwidth]{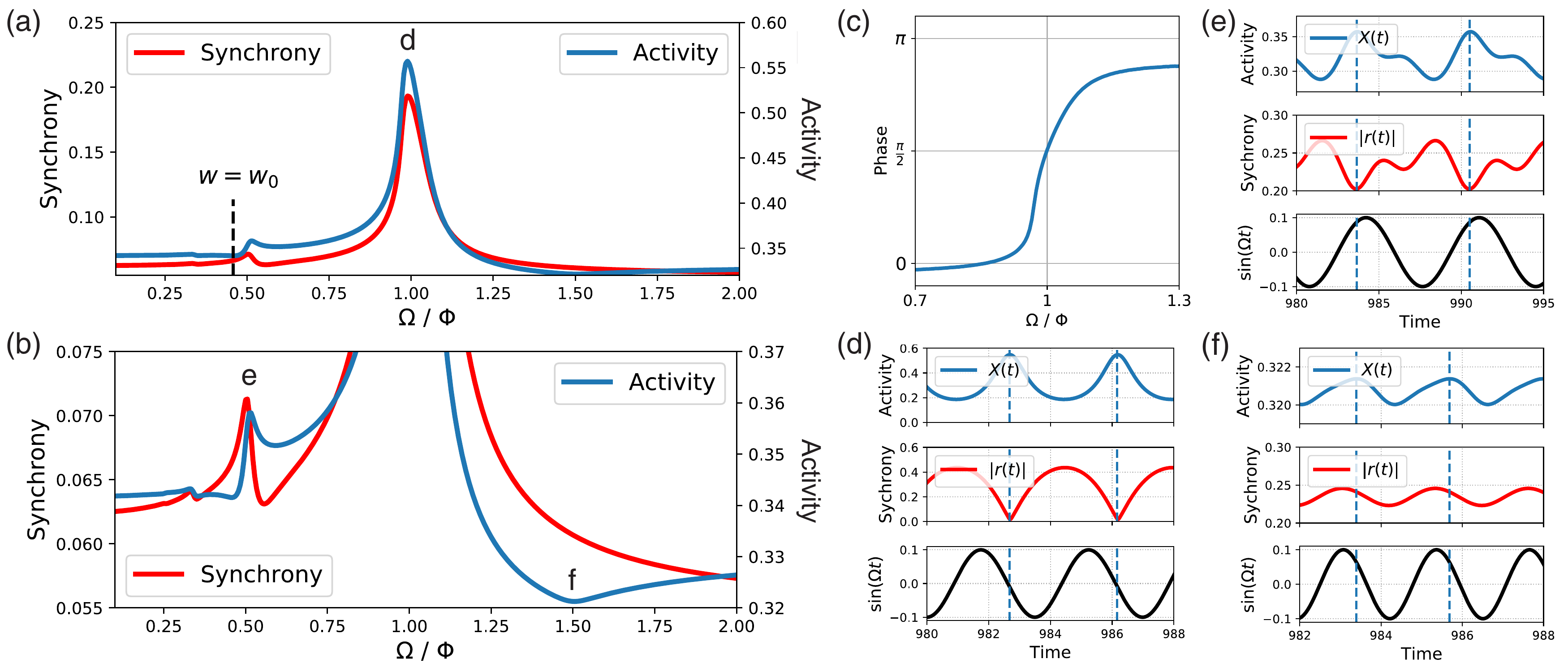} 
  \end{center}
  \caption{Response of the system to the first forcing strategy, an homogeneous forcing with strength $A_0 = 0.1$. The underlying autonomous model has parameters $(w_0,k,\Delta)=(1.1,1.5,0.05)$. (a) Maximum value in a forcing period of $X(t)$ and $\left|r(t)\right|$ for different forcing frequencies $\Omega$ expressed in terms of the underlying autonomous model rotational frequency $\Phi$. (b) Detail of the previous figure. (c) Phase shift between the forcing signal and the systems activity. (d) Time evolution of system's activity and synchrony for the principal maximum $\Omega=\Phi$, (e) for the secondary maximum $\Omega= 0.5 \ \Phi$, and (g) for the minimum $\Omega= 1.5 \ \Phi$. Vertical lines denotes activity's maximums.}
 \label{fig_07}
\end{figure}

As stated previously, this forcing can be thought of as a time modulation of the parameter $w_0$ in the autonomous model: $w_0 \rightarrow \tilde{w_0}(t)= w_0 + A\ sin(\Omega t)$. We will consider a region where this parameter trajectory $\tilde{w_0}(t)$ is completely contained in region $III$ (Figure \ref{fig_04}a) and does not cross any bifurcation plane. In this region the autonomous system has a unique rotational fixed point.

In Figure \ref{fig_07}a, we present the maximum value of Activity and Synchrony, $X(t)$ and $\left|r(t)\right|$, for different forcing frequencies $\Omega$. The behavior is equivalent to that found on resonant systems: there is a principal maximum when the forcing frequency $\Omega$ matches the underlying autonomous model rotational frequency $\Phi$. In this parameter region studied, the system's activity frequency is the same as the forcing one. Moreover, the phase shift between the forcing and the activity has the characteristic shape of a resonant system, being $ \pi / 2$ when both frequencies are equal. The phase shift curve is presented in Figure \ref{fig_07}c. There also exist secondary maximums for the values $\Omega=\Phi/N$ with $N\in \mathbb{N}$, which are appreciable in Figure \ref{fig_07}b. This is a characteristic signature of nonlinear resonances. Additionally, there is a minimum of the system's activity for $\Omega=3\Phi / 2$.


Notice that the system presents a maximum in terms of its activity for the same $\Omega$ value in which it presents a maximum of the synchrony between its units. This means that units need to be more synchronized in order to generate a greater activity. But when inspected temporally, activity and synchrony often present a negative correlation. This is because the synchronicity's maximum takes place when the units get accumulated near $\theta = \pi/2$ due to the critical slowing down of the Saddle-node bifurcation. While the activity's maximum takes place when the units cross $\theta = 3 \pi/2$ and become more desynchronized. Figures \ref{fig_07}c, \ref{fig_07}d and \ref{fig_07}e show the time evolution of the activity and synchrony of the system for cases $\Omega=\Phi$, $\Omega=\Phi/2$ and $\Omega=3\ \Phi/2$ respectively. 

An important observation is that knowing only the intrinsic frequency of the most common units in the system (the ones with $w\approx w_0$) it is not sufficient for predicting anything about the response of the system at a population level. To highlight this fact, the intrinsic frequency of a unit with $w=w_0$ is marked with a dotted line in figure \ref{fig_07}a.

\subsubsection{Heterogeneously forced with continuous distribution}

\begin{figure}[!ht]
  \begin{center}
    \includegraphics[scale=.7]{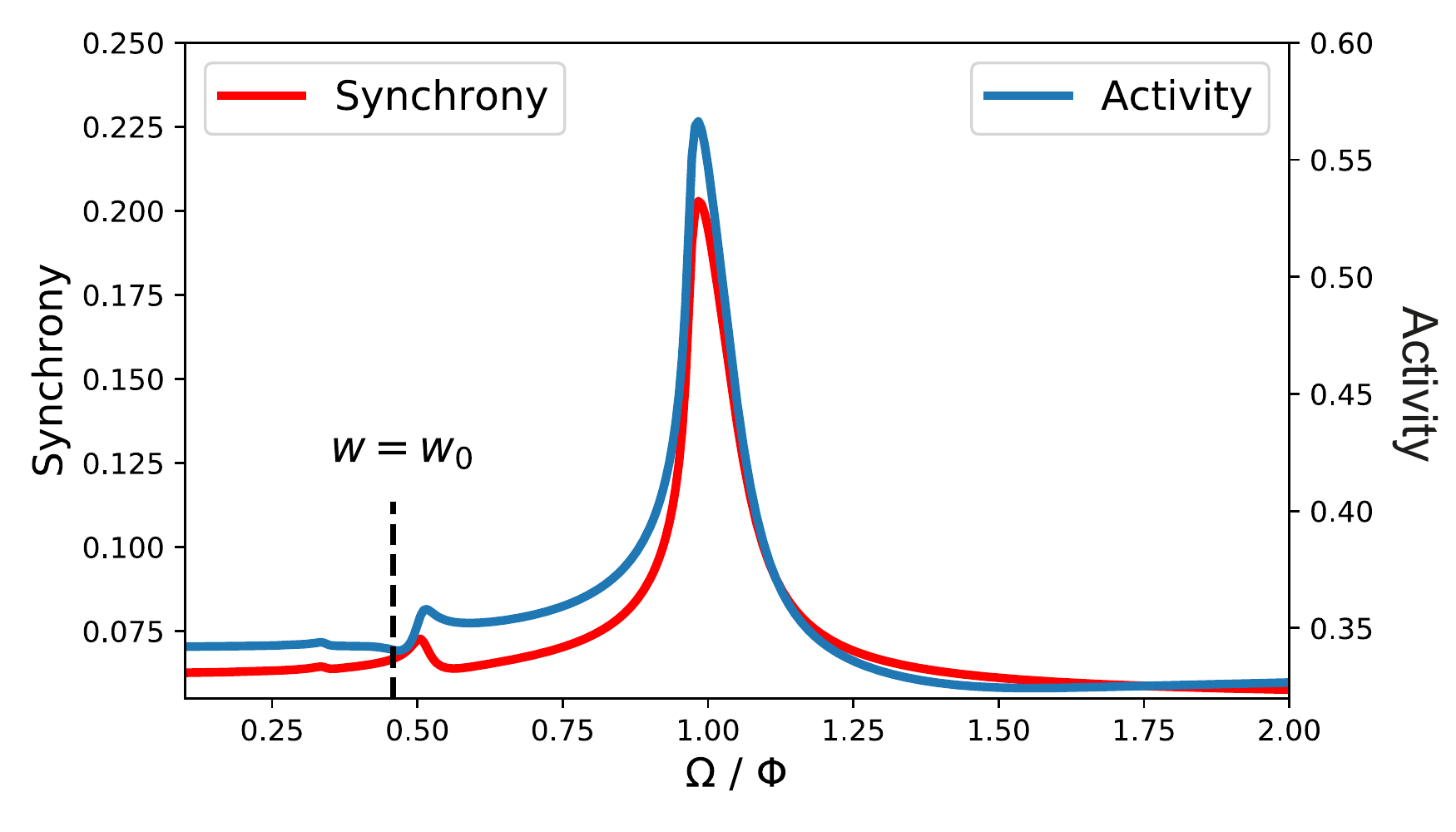} 
  \end{center}
  \caption{Response of the system to the second forcing strategy, an heterogeneous forcing with a continuous $\Gamma_2(A)$ distribution of parameters $A_0 = 0.1$ and $\Delta_A = 0.025$. The underlying autonomous model has parameters $(w_0,k,\Delta)=(1.1,1.5,0.05)$. In the figure, the maximum value in a forcing period of $X(t)$ and $\left|r(t)\right|$ for different forcing frequencies $\Omega$ are presented.}
 \label{fig_08}
\end{figure}

In this second case, the forcing can also be thought of as a time modulation of the parameters in the autonomous model $(w_0,\Delta) \rightarrow (\tilde{w_0}(t),\tilde{\Delta}(t))= (w_0+A\ sin(\Omega t),\Delta+\Delta_A \ sin(\Omega t))$. As in the previous scenario, we are considering a case where the parameters trajectory $(\tilde{w_0}(t),\tilde{\Delta}(t))$ is completely contained in region $III$ (Figure \ref{fig_04}a).

The response of the system to this second forcing strategy, presented in figure \ref{fig_08}, is essentially the same as that found for the first one. So, for this parameter region, introducing a continuous lorentzian shaped heterogeneity in the forcing strength $A$ does not make a significant change in the system's response to that forcing.

\subsubsection{Heterogeneously forced with discrete distribution}

\begin{figure}[!ht]
  \begin{center}
    \includegraphics[width=\textwidth]{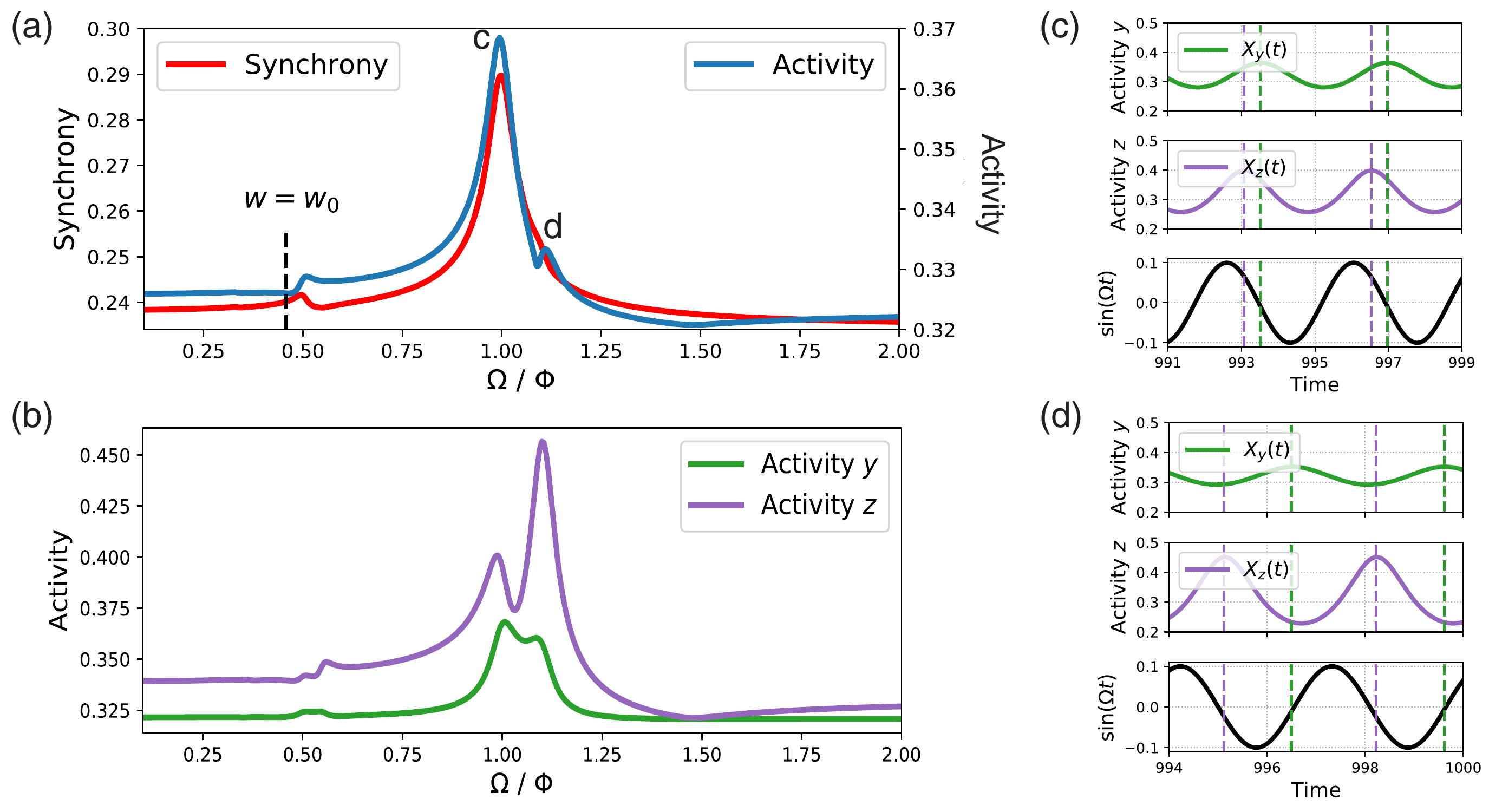} 
  \end{center}
  \caption{Response of the system to the third forcing strategy, an heterogeneous forcing with a discrete $\Gamma_3(A)$ distribution of parameters $A_0 = 0.1$ and $p=0.25$. The underlying autonomous model has parameters $(w_0,k,\Delta)=(1.1,1.5,0.05)$. (a) Maximum value in a forcing period of $X(t)$ and $\left|r(t)\right|$ for different forcing frequencies $\Omega$ expressed in terms of the underlying autonomous model rotational frequency $\Phi$. (b) Maximum value in a forcing period of $X(t)$ for the subpopulation of connected units $z$ and the one of non-connected units $y$.(c) Time evolution of system's activity and synchrony for the principal maximum $\Omega=\Phi$, and (d) for the secondary maximum $\Omega= 1.11 \ \Phi$. Verticals lines denotes activity's maximums of both populations.}
 \label{fig_09}
\end{figure}

For the last forcing strategy, we considered the same scenario as the first case, but this time with only a quarter of the units connected to the external forcing, i.e., with $c=0.25$. The response of the system is presented in Figure \ref{fig_09}a. In this case, a secondary maximum appears for $\Omega \approx 1.11 \Phi$. The location of this new peak does not depends on the proportion of connected units $p$, only it's amplitude does.

Forcing only part of the neurons generates two distinguishable subpopulations, which becomes evident in terms of the model equations (\ref{EcusHeter}). The external input drives the connected subpopulation, $z$, which in turn drives the non-connected one, $y$. And, when driven, subpopulation $y$ also affects on population $z$. In Figures \ref{fig_09}b and \ref{fig_09}c we present the time evolution of the activity for each of the subpopulations.

The system's activity results from the weighted sum of both population's activities, $X_3(t)=(1-p) \ X_y(t)  + p \ X_z(t)$, so it's maximum depends not only on the maximums values of $X_y(t)$ and $X_z(t)$, but also on the phase difference between them. Figure \ref{fig_09}d shows the activity maximum for each population as a function of the forcing frequency. As appreciated from the figure, and also from Figures \ref{fig_09}b and \ref{fig_09}c, the activity maximum of the connected population $z$ is bigger for $\Omega = 1.1 \  \Phi$ than for $\Omega = \Phi$, while the activity maximum of non-connected population $y$ is only slightly smaller. On the other hand, for the total population the activity maximum results significantly bigger for the case $\Omega = \Phi$. This is because the phase difference in this case is such that both subpopulation's activities sum in a constructive way.

As in previous cases, the response of the system presents a resonant behavior when the forcing matches the frequency of the underlying autonomous system. But it is remarkable that, only by changing the number of connected units to the external input, a new relevant frequency appears.

\section{CONCLUSIONS}
\label{sec:5}

For many years, empirical models were the natural choice for mathematicians and physicists to study brain activity. Recently, new analytical tools have been developed, allowing us to construct simple models for brain activity from first principles, i.e. by modeling an heterogeneous population of coupled units. Therefore, there is now a major interest in understanding if the dynamic displayed by this kind of models can reflect some important properties of the system that were missing in the traditional empirical models.

We found that the collective dynamics of a simple ensemble of excitable units can present oscillatory transients in the system's average activity, which can be driven to resonance. We explored different forcing strategies, what allowed us to understand the dynamical origin of resonances in extended systems, as well as how the macroscopic evolution of the system becomes progressively complex as heterogeneities are considered.

\section{FOUNDING}

This work describes research partially funded by National Council of Scientific and Technical Research (CONICET), National Agency of Science and Technology (ANPCyT), University of Buenos Aires (UBA) and National Institute of Health through R01-DC-012859.


\bibliographystyle{ws-ijbc}

\bibliography{Bibliography}

\end{document}